\documentclass{aastex63}%
\usepackage{amsmath}
\usepackage{graphicx}
\usepackage[caption=false]{subfig}
\usepackage{float}
\usepackage{txfonts}
\received{January 15, 2020}
\accepted{March 20, 2020}
\submitjournal{ApJL}

\shorttitle{3D Rad-MHD simulations of starspots}
\shortauthors{Panja et al.}

\begin{document}

   \title{Sunspot simulations: penumbra formation and the fluting instability }


\correspondingauthor{Mayukh Panja}
\email{panja@mps.mpg.de}

\author[0000-0003-0832-4117]{Mayukh Panja}
\affiliation{Max Planck Institute for Solar System Research \\
Justus-Von-Liebig-Weg 3, D-37077, Göttingen, Germany \\}

\author[0000-0001-9474-8447]{Robert Cameron}
\affiliation{Max Planck Institute for Solar System Research \\
 Justus-Von-Liebig-Weg 3, D-37077, Göttingen, Germany \\}

\author[0000-0002-3418-8449]{Sami K. Solanki}
\affiliation{Max Planck Institute for Solar System Research \\
Justus-Von-Liebig-Weg 3, D-37077, Göttingen, Germany \\}
 \affiliation{School of Space Research, Kyung Hee University \\
Yongin, Gyenoggu-Do, 446-701,Korea\\}

  \begin{abstract}
  The fluting instability has been suggested as the driver of the subsurface structure of sunspot flux tubes. We conducted a series of numerical experiments where we used flux tubes with different initial curvatures to study the effect of the fluting instability on the subsurface structure of spots. We used the MURaM code, which has previously been used to simulate complete sunspots, to first compute four sunspots in the slab geometry and then two complete circular spots of opposite polarities. We find that the curvature of a flux tube indeed determines the degree of fluting the flux tube will undergo - the more curved a flux tube is, the more fluted it becomes. In addition, sunspots with strong curvature have strong horizontal fields at the surface and therefore readily form penumbral filaments. The fluted sunspots eventually break up from below, with lightbridges appearing at the surface several hours after fluting commences. 

  \end{abstract}


%

\section{Introduction}

It is not known what the magnetic field associated with sunspots looks like underneath the solar surface. 
\cite{Cowling1946} proposed that a sunspot extends below the surface as a magnetic flux tube - field lines bound tightly together in a single monolithic column resisting deformation against pressure from the surrounding gas.  However, the sharp vertical gradient in the ambient gas pressure at the surface necessitates that the magnetic field lines fan out rapidly. This would make a flux tube highly concave near the surface, and therefore susceptible to the fluting instability. This prompted \cite{1979Parker} to suggest an alternative configuration in which the field underneath the surface may be structured -  a sunspot, in this view, is a cluster of numerous small flux tubes that are held together by a converging flow below a certain depth.  However, \cite{MeyerWeiss} had used a vacuum model of a flux tube to study the stability of spots against the fluting instability, and concluded that spots should not break up into smaller flux tubes up to a depth of 5 Mm. \cite{1981Spruit} built on the work of \cite{MeyerWeiss} and constructed a cluster model of a sunspot which is similar to a tethered balloon model (see Figure 1 of \cite{1981Spruit}) - the tube remains coherent up to a certain depth, beyond which it is fragmented into small individual flux tubes that are tied together at the base of the convection zone. It differed from \cite{1979Parker}, in that the tying of the flux tube to the base of the convection zone removed the necessity
of a converging flow to explain the stability of sunspots. For a discussion on the merits and demerits of both the monolithic and cluster models, see Chapter 1 of \cite{Thomas_Weiss1992}. 

The fact that penumbral filaments often invade a spot's umbra and fragment it (  \cite{Spot_frag12,umbra_divide_2018}), suggest that the fluting instability might play a role in determining the subsurface structure of spots and therefore, by extension, their appearance on the surface. However, the probing of sunspot subsurface structure using helioseismic techniques has not been able to distinguish between the cluster and monolithic models \citep{Moradi2010SoPh}. Existing MHD simulations of complete sunspots, \citep{rempel09b,Rempel_2011_subsurface,Rempel_2011b} using the radiative-MHD code MURaM \citep{Vogler2005,rempel09a}, correspond to the monolithic model. \cite{Rempel_2011_subsurface} specifically addressed the question of whether a sunspot is monolithic or cluster-like underneath the surface and concluded that sunspots are closer to the monolithic model, but can become highly fragmented in its decay stage.  However, these models have field lines that are too vertical near the spot periphery to form penumbral filaments naturally. This is overcome by increasing  the horizontal field strength at the top boundary by a factor of two compared to a potential field configuration, and the extent of the penumbra is solely determined by the magnetic top boundary condition \citep{Rempel_2012}. Recently, \cite{2020A&A..Jurack} presented a sunspot simulation with a decent sized penumbra without modifying the top boundary, by using a strongly compressed flux tube at the lower boundary. Their penumbra, however, is dominated by the counter-Evershed flow. Also, their  umbral field strength is higher than what is observed.



In this paper, we conduct numerical experiments using the MURaM code to investigate the susceptibility of flux tubes to the fluting instability by varying the initial magnetic field structure. 
We focus on the question - would sunspots with field lines inclined strongly enough to form penumbral filaments, result in flux tubes that become highly fluted under the surface? To this end, we constructed initial sunspot flux tube configurations where the field lines are curved near the surface, such that they form penumbral filaments without having to change the top boundary condition, and become close to vertical below a certain depth. 

We describe our simulation setups and  detailed descriptions of our initial conditions for our magnetic flux tubes in Section \ref{sec:simset}. We conducted four runs in the computationally inexpensive slab geometry, where we systematically varied the radius of curvature ($R_\mathrm{c}$) to check if we can control the degree of fluting. Then we computed two complete circular spots of opposite polarities in a shallow computational domain. We present our results in Section \ref{sec:results} and discuss the implications of our results in Section \ref{sec:diss}. 

\section{Simulation Setup} \label{sec:simset}
We used the MURaM radiative MHD code for our simulations. For our four slab geometry runs, we chose simulation boxes with  dimensions of 36 Mm ($x$) $\times$ 6 Mm ($y$) $\times$ 10.3 Mm ($z$) and resolutions of 48 km $\times$ 48 km $\times$ 25.8 km. We conducted a further run where we computed complete circular spots of opposite polarities. We used a relatively shallow domain with a vertical extent of 6 Mm which had a resolution of 20 km. The  horizontal extents of this run were 72 Mm $\times$ 36 Mm with a resolution of 48 km in both directions.  All of our boxes were periodic in the horizontal directions and the upper boundaries were kept open to plasma flows. When our hydrodynamic runs achieved thermal equilibrium, we introduced magnetic flux tubes in the simulation domains. We initialized our magnetic runs by damping all three components of the velocity field by a factor of $(1+ (|B|^2/B_\mathrm{c}^{2}))$, where $B_\mathrm{c}^{2}$ = 80000 $G^{2}$. We do this only at the timestep where our magnetic flux tubes are introduced, and thereafter we let the convective flow field develop naturally. In the following paragraphs we have described the initial structure of these flux tubes.

\begin{figure*}
    \centering
   	\hspace{-3.0cm}\includegraphics[scale=0.15]{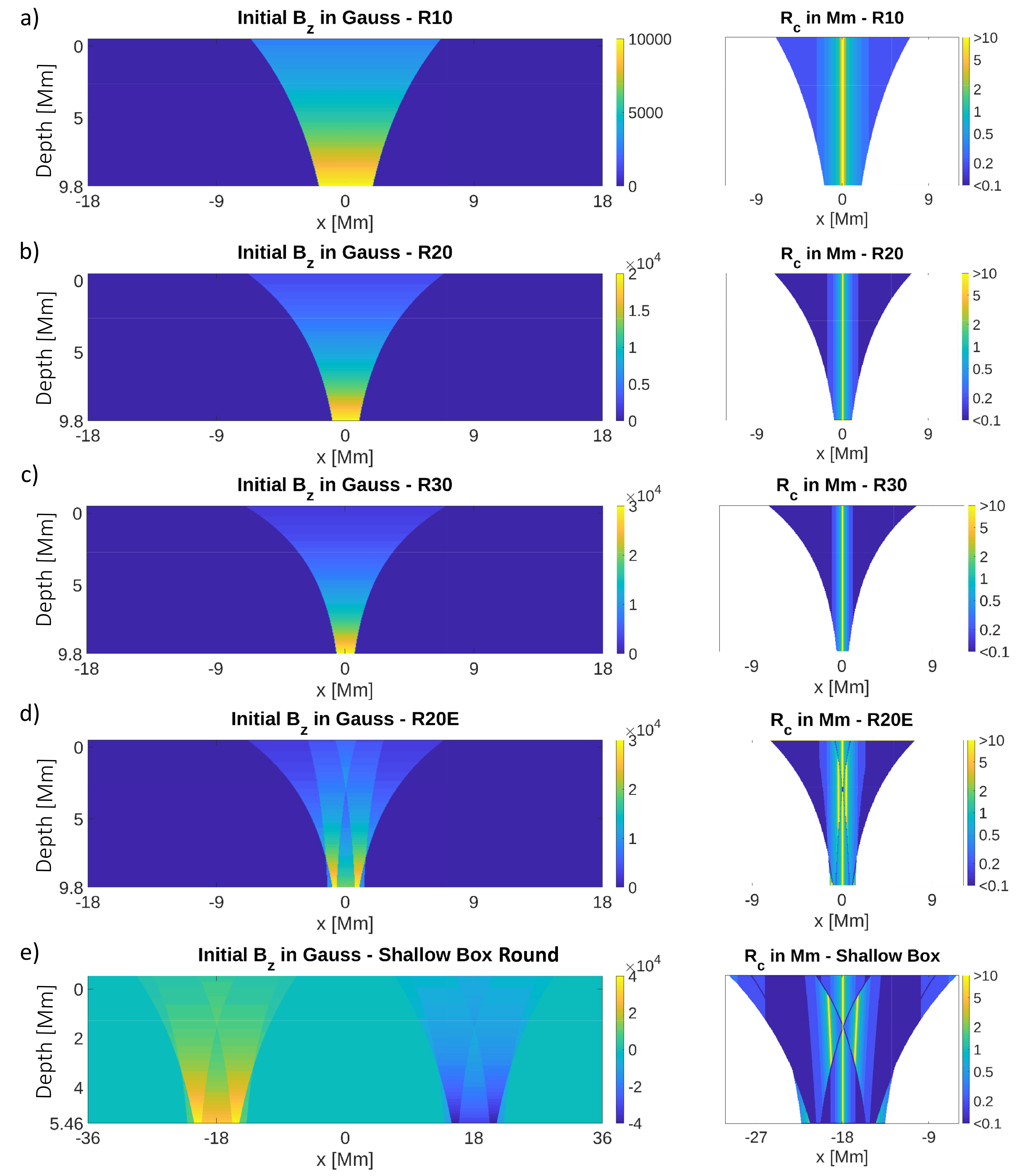}
    \caption{Panels a-d: Vertical slices of B$_\mathrm{z}$ (x-z plane, here Depth = $h_\mathrm{phot}$ - z) used as initial conditions for the slab geometry runs, with their corresponding $R_\mathrm{c}$ shown on the right. The initial magnetic conditions for the slab geometry runs were invariant in the y direction. Panel e:   Vertical slice of B$_\mathrm{z}$ used as initial condition for the circular spot simulation. The circular spot had an axisymmetric initial condition, with its corresponding $R_\mathrm{c}$ shown on the right. }
    \label{fig:slabinit}
\end{figure*}   

\subsection{Slab Geometry Runs}
As discussed, our initial conditions are designed to serve two purposes - 1) they should result in the formation of penumbral filaments, and 2) have a small radius of curvature ($R_\mathrm{c}$) that induces the fluting instability. Since these are numerical experiments, we are free to choose the initial conditions to achieve these goals. 
For our slab geometry runs, we define the three components of the magnetic field inside the initial flux tubes, in conformity with the $\Vec{\nabla} \cdot \Vec{B} = 0$ constraint, as follows: 

\begin{align}
\label{eqn:1}
 B_{z} & = f(z), \nonumber \\ 
 B_{x} & = -xf'(z), \nonumber \\
 B_{y} & = 0, 
\end{align}

where,
\begin{equation}
f(z) =B_{bot} \exp{\frac{-z}{\sigma}}. 
\label{eqn:fz}
\end{equation}
At $z=0$ (the lower boundary) we set $B_{z}$ to $B_{bot}$ and at $z = h_{opt}$ (optical surface) , we set $B_{z}$ to $B_{opt}$. $B_{opt}$ and $B_{bot}$ are parameters that we are free to choose. Using these constraints and eqn. \ref{eqn:fz} we can express $\sigma$
as, 

\begin{equation}
\sigma = h_{phot}/\log(\frac{B_{bot}}{B_{opt}}).     
\end{equation}  

Keeping $B_\mathrm{opt}$ at 3000 Gauss, we conducted three runs with $B_\mathrm{bot}$ as 10000, 20000 and 30000 Gauss. We labeled these runs as R10, R20 and R30 respectively. The top 3 panels of Figure \ref{fig:slabinit} depict the initial $B_\mathrm{z}$ and their corresponding $R_\mathrm{c}$, for these runs.

In order to quantify $R_\mathrm{c}$ we have to first calculate the curvature vector ($\Vec{\kappa}$), which is given by:
\begin{equation}
 \Vec{\kappa} = \Vec{b} \cdot \Vec{\nabla} \Vec{b}
\end{equation}

where,

\begin{equation}
 \Vec{b} = \frac{\Vec{B}}{|\Vec{B}|}
\end{equation}

The inverse of the magnitude of $\Vec{\kappa}$ $(\frac{1}{|\Vec{\kappa}|})$ at any point, gives the local $R_\mathrm{c}$. We have plotted the corresponding $R_\mathrm{c}$ of our initial magnetic fields in the right hand column of Figure \ref{fig:slabinit}. In all of the cases, $R_\mathrm{c}$ is very high at the centre, implying near vertical fields, while at the edges the fieldlines are significantly curved. Clearly, the fieldlines become more curved as we progress from  R10 to R30 (panels a - c). Note how the brighter band in the centre, becomes narrower from R10 to R30. One can predict thus, fluid elements can penetrate the furthest into the flux tube of R30 before meeting any resistance from strong vertical fields. A side effect of decreasing $R_\mathrm{c}$ simply by continuously increasing the field strength at the lower boundary is that it keeps making the flux tube narrower at its base. We, therefore, carried out another experiment where we tried out a different initial condition. We superimposed two additional flux tubes on either side of the main flux tube used in R20, as shown in  panel d of Figure \ref{fig:slabinit}. We did this because - 1) the enhanced field strength at the edges, close to the lower boundary, would help keep the flux tube coherent at the base of the simulation box (note that this run has the highest $R_\mathrm{c}$ at the base) 2) the additional magnetic pressure around the centre of the flux tube, near the surface, would help the fieldlines fan out more and become even more inclined once the flux tube achieves pressure equilibrium, facilitating penumbral filament formation. We labeled this run R20E. Note that due to the superposed smaller tubes, the initial field strength at the base of the computational domain in this run locally reaches 30 kG at the edges.

\subsection{Round spots}

For our shallow round spot simulation we use an initial condition, which has a  vertical cut similar to the vertical cut of the initial condition used in R20E. Two flux sheets were superimposed on either side of the main flux sheet and this was rotated  axisymmetrically, while ensuring that $\nabla\cdot B=0$. A vertical cut of the initial condition through the centre of the simulation box is shown in panel e of Figure \ref{fig:slabinit}. 

\subsection{Boundary Condition for the magnetic field}
In the shallow sunspot simulation presented in \cite{Rempel_2011_subsurface},  a lower boundary open to plasma flows inside the magnetic flux tube caused the sunspot to disintegrate completely within 6 hours. In our simulations, for all of the runs, we  set all velocities to zero at the lower boundary for $|B|> 1000$ Gauss. This allows us to study the effects of the fluting instability with minimal interference from the lower boundary. At  the upper boundary the magnetic field was made to have a potential field configuration.


\section{Results} \label{sec:results}

\begin{figure*}
   \centering
   \includegraphics[scale=0.55]{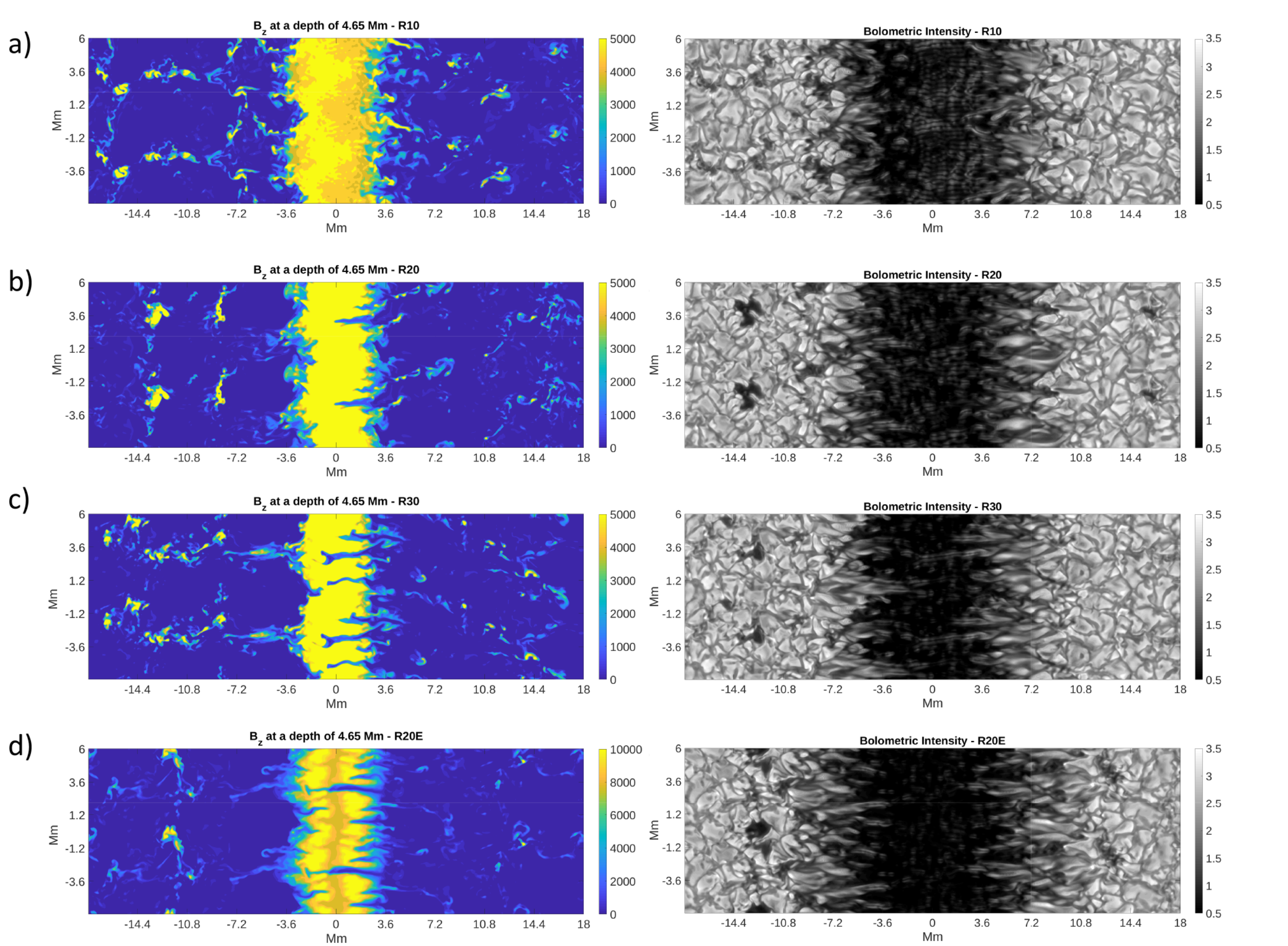}
    \caption{Left Panel: Horizontal cuts of B$_\mathrm{z}$ of the slab geometry runs - R10, R20, R30, R20E, in Gauss at a depth of 4.65 Mm after 8 hours of solar runtime. Right Panel: The corresponding bolometric intensity maps in units of 10$^{10}$ erg cm$^{-2}$ ster$^{-1}$ s$^{-1}$. The images have been repeated twice in the y-direction.
               }
        \label{Fig:slabintenmidbox}
    \end{figure*}    

\begin{figure*}
   \centering
   \includegraphics[scale=0.3]{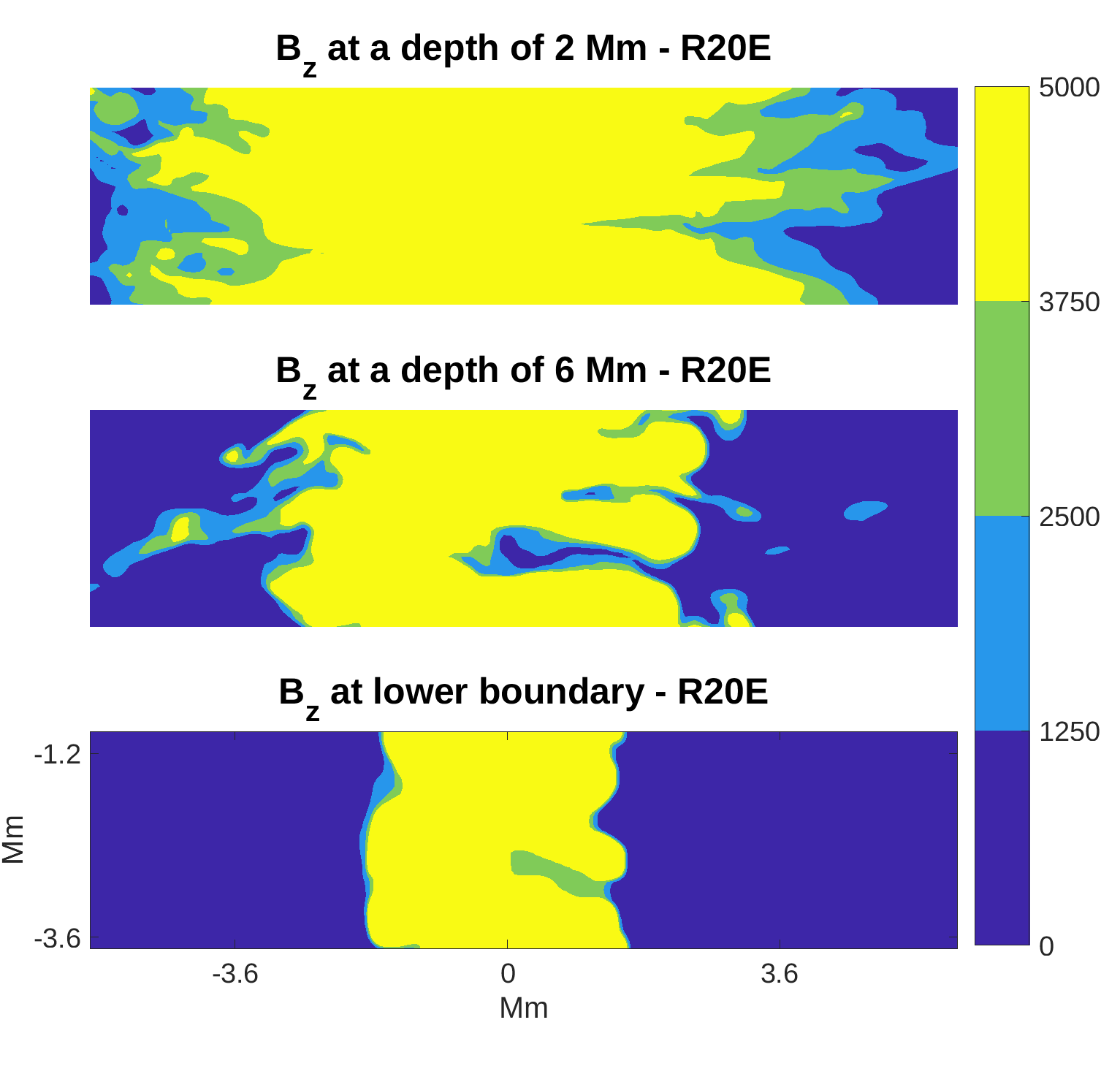}
    \caption{Zoomed in horizontal cuts of B$_\mathrm{z}$ (Gauss) of the run R20E at different depths below the photosphere. We have intentionally chosen only 4 contour levels to draw attention to the tongue like weak field regions at the edge of the flux tube  caused by fluid penetrating from outside.
               }
        \label{Fig:bz_horcuts_r20e}
\end{figure*}

\begin{figure*}
    \centering
    \hspace{-1.5cm}\includegraphics[scale=0.45]{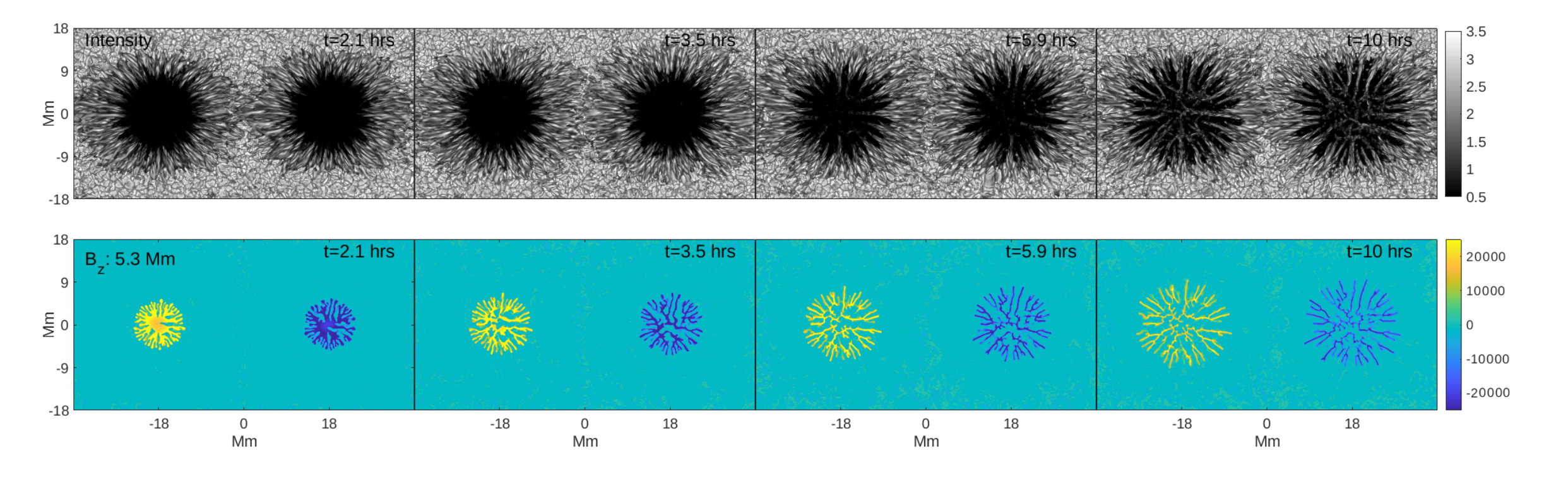}
    \caption{Temporal evolution of the circular spot simulation showing the advancement of the fluting instability. The top panel shows the emerging bolometric intensity in units of 10$^{10}$ erg cm$^{-2}$ ster$^{-1}$ s$^{-1}$ at different stages of the evolution. The lower panel shows horizontal cuts of B$_\mathrm{z}$ in Gauss at a depth of 5.3 Mm below the photosphere. }
    \label{fig:time_ev}
\end{figure*} 

\subsection{Slab Geometry Runs}

The left panel of Figure \ref{Fig:slabintenmidbox} shows horizontal cuts of B$_\mathrm{z}$ at a depth of 4.65 Mm below the visible surface. It is clear that both the number of filament-like intrusions of the surrounding plasma and the lengths of such intrusions, increase as we increase the curvature of the initial flux tubes, as seen in the results of R10, R20 and R30. In all of the runs, the instability originates close to the middle of the box, where the curvature is maximum, and propagates both upwards and downwards. Some of these intrusions eventually manifest themselves at the surface in the intensity images as long penumbral filaments with thin dark cores (see right panel of Figure \ref{Fig:slabintenmidbox}). The purpose of the runs in the slab geometry was to vary R$_\mathrm{c}$ and see if it results in different amounts of fluting. Our results confirm that R$_\mathrm{c}$ indeed controls the degree of fluting. 
 
The run R20E exhibits properties that lie  between R20 and R30 - the intrusions are plentiful but only a couple of them manage to reach the centre of the flux tube. At the surface, it develops the most expansive penumbra among the four cases, while having an umbra that is not distorted by intruding filaments. This indicates that our numerical experiment of superimposing two additional flux tubes achieved its intended purpose. This prompted us to choose the initial condition for the next circular spot simulation such that its vertical slice is similar to run R20E.  

A side effect of the higher field strengths in the lower boundary is that the runs R30 and R20E have comparatively cleaner umbrae with fewer umbral dots.

In Figure \ref{Fig:bz_horcuts_r20e} we have plotted horizontal cuts of B$_\mathrm{z}$ at different depths of the R20E run. We have zoomed in on only a part of the flux tube so that we can investigate individual filaments. We have chosen only 4 contour levels so that we can easily discern the penetrating tongues of the external fluid. Notice that the tongue-like weak field regions are the most prominent at a depth of 6 Mm,  while both at the lower boundary and at a depth of 2 Mm only traces of the intrusions have appeared.  It is clear that the fluting originates near the middle of the box and propagates both upwards and downwards through diffusive processes and pressure differences generated by the penetrating plasma. This also demonstrates that the fluting is not merely a boundary effect.

\subsection{Round spots}

For our circular spot simulation, we used initial conditions that are similar to the one used in run R20E. Close to the surface, the initial flux tube had strong vertical fields near the centre, while below a certain depth the field strength at the edges of the flux tube was enhanced. We have plotted in Figure \ref{fig:time_ev} (part a) the evolution of the circular spot simulation in the shallow box. The top panel shows a series of intensity images at different stages of the evolution, while the bottom panel shows the corresponding horizontal cuts of  B$_\mathrm{z}$ at a depth of 5.3 Mm. As seen in the intensity image panel, the inclined fields near the surface and the presence of opposite polarities result in the formation of penumbral structures of considerable extent in both the positive and negative spots 2 hours into the run. By this time, the corresponding flux tubes already show a very high degree of fluting. In the subsequent time frames, the flux tubes get more and more distorted and 6 hours into the simulation they are no longer coherent and break up into disconnected fragments. The instability propagates upwards and we see the head of the filaments gradually penetrating the umbral regions. The last snapshot has been taken 10 hours into the run and by this time the umbra in the intensity image is completely covered with protruding filaments whose heads have migrated all the way to the center. The corresponding horizontal cut shows that the  flux tubes are completely distorted and they are both reminiscent of the spaghetti-like structure  hypothesized by \cite{1979Parker}. In our simulations, we see multiple flux sheets form, some of them loosely connected. It is important to note that in addition to being fluted the flux tubes are also continuously pulled apart by convection and we see the circumferences of both the tubes expanding with time. This accelerates the breaking up of the flux tubes into individual components which in turn facilitates the filaments at the surface to penetrate further into the umbrae. This is in agreement with \cite{1979Parker} who suggested that in order to prevent a fluted flux tube from being completely pulled apart there must be a converging flow that holds the different parts together and in the absence of a converging flow in our simulations, the flux tubes simply break up. It is important to bear in mind that we had set all velocities at points with $|B|> 1000$ G at the lower boundary to zero. However, the magnetic field  at the lower boundary  can  still be transported by the external flow field and be weakened by filamentary intrusions from above, mediated by diffusive processes.

    

\begin{figure*}
   \centering
	\hspace*{-1.0cm}\includegraphics[scale=0.06]{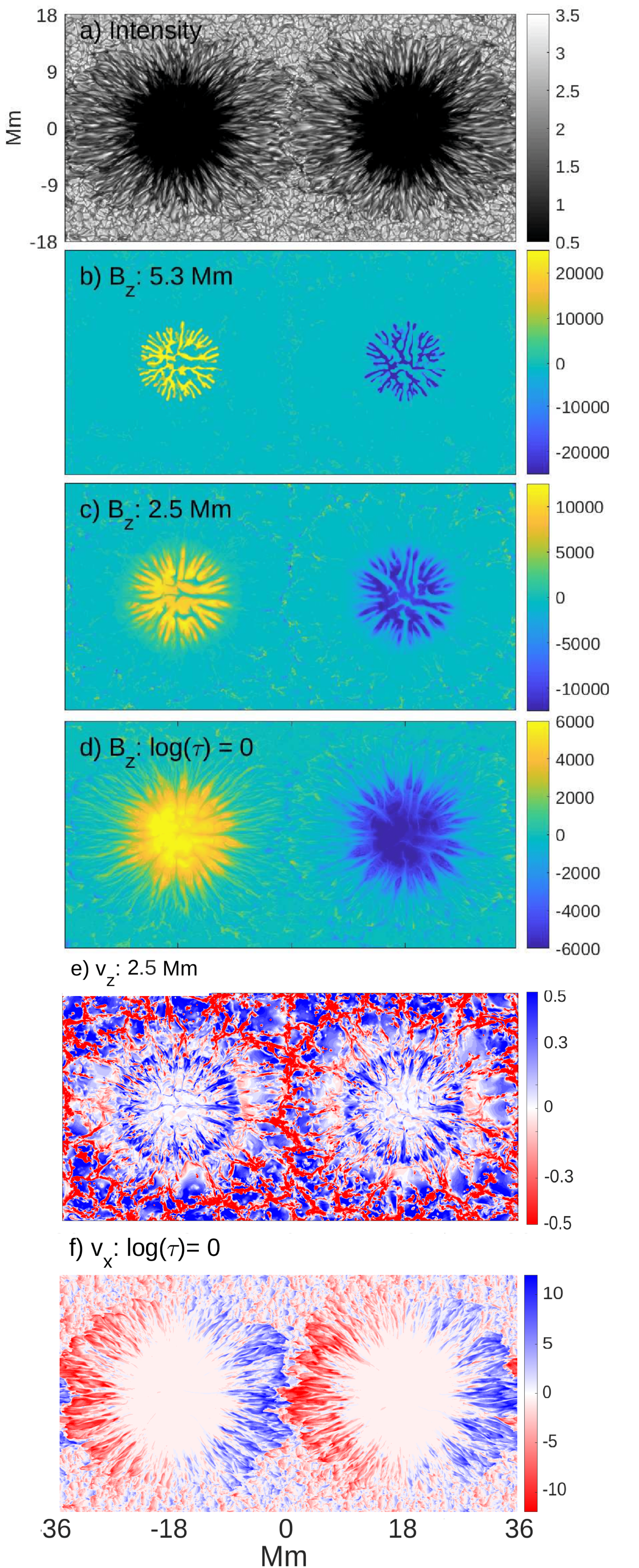}
   
    \caption{Snapshot of the circular spot simulation after 3.5 hours of solar runtime with the bolometric intensity image in  the top panel (a), horizontal cuts of B$_\mathrm{z}$  at two different depths (b-c) and at the $\tau$=1 surface (d).  Panel \textit{e} shows the vertical velocity profile at a depth of 2.5 Mm and  panel \textit{f} plots v$_\mathrm{x}$ at the $\tau$=1 surface. The intensity map is in units of 10$^{10}$ erg cm$^{-2}$ ster$^{-1}$ s$^{-1}$, B$_\mathrm{z}$ is in Gauss, and the velocities are  in units of km/s.
               }
        \label{Fig:round_diffdepths}
    \end{figure*}

In Figure \ref{Fig:round_diffdepths}, we have presented after 3.5 solar hours the bolometric intensity image (panel a),  horizontal cuts of the magnetic field at different depths (panels b-d), the vertical velocity profile at a depth of 2.5 Mm (panel e) and the velocity along the x direction at the $\tau=1$ surface (panel f). At a depth of 5.3 Mm, the flux tubes are almost completely shredded after 3.5 hours of runtime. The instability, in this case, had  originated closer to the lower boundary and propagated upwards as is evidenced by the decreasing severity of the fluting at depths of 2.5 Mm and the $\tau$=1 surface. In panel e, we have plotted v$_\mathrm{z}$ at a depth of 2.5 Mm. We find that in the areas that correspond to the penetrating fluid at the edge of the flux tube, there is a systematic upflow. These upflows eventually help the intrusions manifest at the surface as lightbridges. At the centre of the flux tube v$_\mathrm{z}$ becomes negligible. A noticeable feature in  the intensity image is the extent of the penumbra. We have achieved an umbra:penumbra area ratio of around 1:4 which is in the range of what is observed on the Sun \citep{solanki_review}. This is a significant result since sunspot simulations typically use the upper boundary  to achieve respectable penumbral proportions. In contrast to \cite{2020A&A..Jurack}, who also used the subsurface structure of the sunspot to produce a penumbra, we obtain Evershed flows that have the correct orientation (panel f). We, however, also obtain umbral field strengths that are higher than what is typically observed, like \cite{2020A&A..Jurack}. The periodicity of the horizontal boundaries makes the penumbra slightly asymmetric. Thus it is more elongated in the x-direction, where the opposite polarities meet. 


\section{Conclusion} \label{sec:diss}
We have simulated complete sunspots that naturally form penumbral filaments and have further demonstrated that sunspots with highly curved flux tubes may have subsurface structures which are close to the cluster model proposed by \cite{1979Parker}.  Our simulations lead us to make the following conclusions about the nature of sunspots -

1) It is quite clear that the initial subsurface structure plays an important role in the formation of penumbral filaments and the stability of sunspots. Highly curved flux tubes are indeed vulnerable to the fluting instability, as had been speculated by many authors before.  Our experiments in the slab geometry where we systematically varied the curvature of the initial flux tube confirm that the intrusions of plasma into the flux tubes are indeed due to the fluting instability and we could control the degree of fluting to some extent by continuously decreasing the radius of curvature. 

2) Our circular spot simulation has strong horizontal fields and consequently develops extended penumbral filaments that harbour the Evershed flow.

3) Our simulations suggest that even sunspots with little structuring at the surface might  already be highly fluted underneath and eventually the subsurface structuring is manifested at the surface through penumbral filaments encroaching into the umbra. The  nearly field-free material typical of such intruding filaments reach down 5 Mm or more in our simulations. Whether sunspots anchored deep in the convection zone can keep the spaghetti-like structure from being torn apart, as predicted by \cite{1981Spruit}, remains an open question. Sunspot simulations that cover the full convection zone, such as the one by \cite{2020MNRAS..Hotta}, can be used to answer this question.

\acknowledgements
This project has received funding from the European Research Council (ERC) under the European Union’s Horizon 2020 research and innovation programme (grant agreement No. 695075) and has been supported by the BK21 plus program through the National Research Foundation (NRF) funded by the Ministry of Education of Korea. MP acknowledges support by the International Max-Planck Research School (IMPRS) for Solar System Science at the University of Göttingen.  RHC acknowledges partial support from the ERC synergy grant WHOLESUN 2018.  The simulations have been carried out on supercomputers at GWDG and on the Max Planck supercomputer at RZG in Garching.

\bibliography{ms.bbl}{}
\bibliographystyle{aasjournal.bst}

\end{document}